\documentclass[amsmath,amssymb]{revtex4}

\usepackage{bm}
\usepackage{color}
\usepackage{dcolumn}
\usepackage{epsfig}
\usepackage{graphicx,rotating}
\usepackage{longtable}
\usepackage{mhchem}
\usepackage{gensymb}
\usepackage{multirow}
\usepackage{balance}

\usepackage{comment}
\usepackage{physics}
\usepackage[caption=false]{subfig}

\newcolumntype{C}[1]{>{\centering\arraybackslash}m{#1}} 

\newlength{\twosubht}
\newsavebox{\twosubbox}

\begin{document}

\title{Exploring epitaxial structures for electrically pumped perovskite lasers: a study of CsPb(Br,I)$_3$ based on the \textit{ab initio} Bethe-Salpeter equation}

\author{Ma\l{}gorzata Wierzbowska}
\affiliation{nstitute of High Pressure Physics, Polish Academy of Sciences}
\author{Juan J. Mel\'endez}
\affiliation{Department of Physics, University of Extremadura. Avda. de Elvas, s/n, 06006, Badajoz, Spain}
\affiliation{Institute for Advanced Scientific Computing of Extremadura, Badajoz, Spain}

\date{\today}

\begin{abstract}
Halide perovskites are widely used as components of electronic and optoelectronic devices such as solar cells, light-emitting diodes (LEDs), optically pumped lasers, field-effect transistors, photodetectors, and $\gamma$-detectors. Despite this wide range of applications, the construction of an electrically pumped perovskite laser remains challenging. In this paper, we numerically justify that mixing two perovskite compounds with different halide elements can lead to optical properties suitable for electrical pumping. As a reference, the chosen model material was CsPbBr$_3$, whose performance as a part of lasers has been widely recognised, with some Br atoms substituted by I at specific sites. In particular, a strong enhancement of the low-energy absorption peaks has been obtained using the \textit{ab initio} Bethe-Salpeter equation. Based on these results, we propose specific architectures of ordered doping that could be realised by epitaxial growth. Efficient light emission from the bottom of the conduction band is expected.
\end{abstract}

\maketitle

\section{Introduction}

The rapid development of epitaxial growth techniques has made it possible to obtain very high quality crystals with atomically tailored structures. These crystals have the desired properties for use as components in optoelectronic devices. In the context of this paper, the first epitaxial growth of halide perovskites with the formula ABX$_3$ (where A is typically a small organic molecule or an inorganic monovalent cation A$^{1+}$, B is a divalent cation such as Pb$^{2+}$, Sn$^{2+}$ or Ge$^{2+}$, and X is a halide anion Cl$^{1-}$, Br$^{1-}$ or I$^{1-}$) was achieved in 2017 \cite{2017}. Since then, halide perovskites have been grown by many epitaxial techniques \cite{E1} including vapour phase deposition \cite{E2}, electrodeposition \cite{E3}, spin coating \cite{S1} and ion exchange \cite{I1}. 1D structures resembling nanowires have been grown on the steps of the substrate surface \cite{QNW}, while 0D quantum dot-like structures have been obtained by the core-shell epitaxy technique \cite{E4}.

In general, epitaxial growth is inevitably related to the lattice mismatch between adjacent layers. In this sense, the lattice constants of halide perovskites are highly dependent on the type of halide. The iodide compounds have the shortest B-X bonds, while the chlorides have the longest ones, with those of the bromides in between. The corresponding sizes of the fundamental band gaps follow the reverse order of the lattice constants. To reduce the stresses in each deposited layer during epitaxy, the material with the larger lattice constant is grown on top of the compound with the shorter one. This is a common  solution used to produce optoelectronic devices based on wurtzite III-V nitrides. For example, the bond lengths increase in the series of AlN, GaN and InN, and the fundamental band gaps decrease accordingly \cite{NO}. Thus, the common deposition order is InGaN for pure GaN \cite{N1}. The growth of GaN on AlGaN would be desirable but less convenient for doping purposes since only high temperature and high pressure conditions are useful for these crystals \cite{N2, N3}. In contrast, halide perovskites can be grown easily and cheaply using alternative techniques \cite{E1}, which produce layered compounds that are even more stable than the bulk crystals. This has been achieved, for example, by depositing Cl-rich layers on Br-rich substrates with posterior relaxation by high temperature annealing \cite{TT}. In this work we show that doping the shortest lattice constant (and larger bandgap) halide perovskite by depositing one-atom thick interlayers of the largest lattice constant one leads to particular optical properties that could potentially enable the construction of electrically pumped lasers.

The first optoelectronic devices realised with halide perovskites were solar cells \cite{SC-1}, whose efficiency has recently reached 25\% \cite{SC-2}, or even more than 30\% in a hybrid tandem configuration \cite{SC-3}. Although the working principle of a photodetector is similar to that of a solar cell \cite{PH}, it is surprising that X-ray and $\gamma$-detectors have also been reported \cite{gamma-detector}. Besides, field effect transistors \cite{TR-1} and neuromorphic recognition devices \cite{Neu} based on halide perovskites also exist. Regarding lasers, the first optically pumped perovskite lasers were reported in 2014 \cite{L1, L2}. A year later, the first amplified spontaneous emission and lasing with a low pump threshold of 5 $\pm$ 2 $\mu$J cm$^{-2}$ \cite{ASE} and the first perovskite LEDs operating at room-temperature were reported\cite{LED}. Since then, many reviews on lasers and LEDs have been published \cite{R1,R2,R3,R4,R5,R6,R7,R8,R9,R10}. In terms of laser operation, both continuous wave \cite{CW} and ultrashort pulse \cite{SP} lasers are available, and polariton lasers have also been reported \cite{UW}. As for the cavity types, the cavity-free (whispering gallery mode) \cite{WGM}, random \cite{Ran}, microcavity \cite{Micro} and Fabry-Perot \cite{F-P} lasers have been constructed. In terms of dimensionality, perovskite lasers can be divided into many categories, including nanocrystals \cite{cryst}, quasi-2D \cite{q2D}, thin films \cite{film}, nano- and microplates \cite{NP, MP}, nanowires and nanoribbons \cite{NW, NR}, quantum dots embedded in silica spheres \cite{QD1} or in glass tubes \cite{QD2}, colloids \cite{R6} and arrays \cite{arrays}, among others. In spite of the above-mentioned successes in optical pumping, the construction of electrically pumped perovskite lasers is still the subject of scientific research
. Many experimental attempts have led to the emergence of several lines of research to achieve this goal, usually based on the improvement of the parameters characterising LEDs \cite{x1,x2,x3,x4,x5,x6}. In this theoretical work, we focus on the atomistic design of optically active materials with tailored excitonic properties.

\begin{figure}[!ht]
    \centering
    \includegraphics[width=\textwidth]{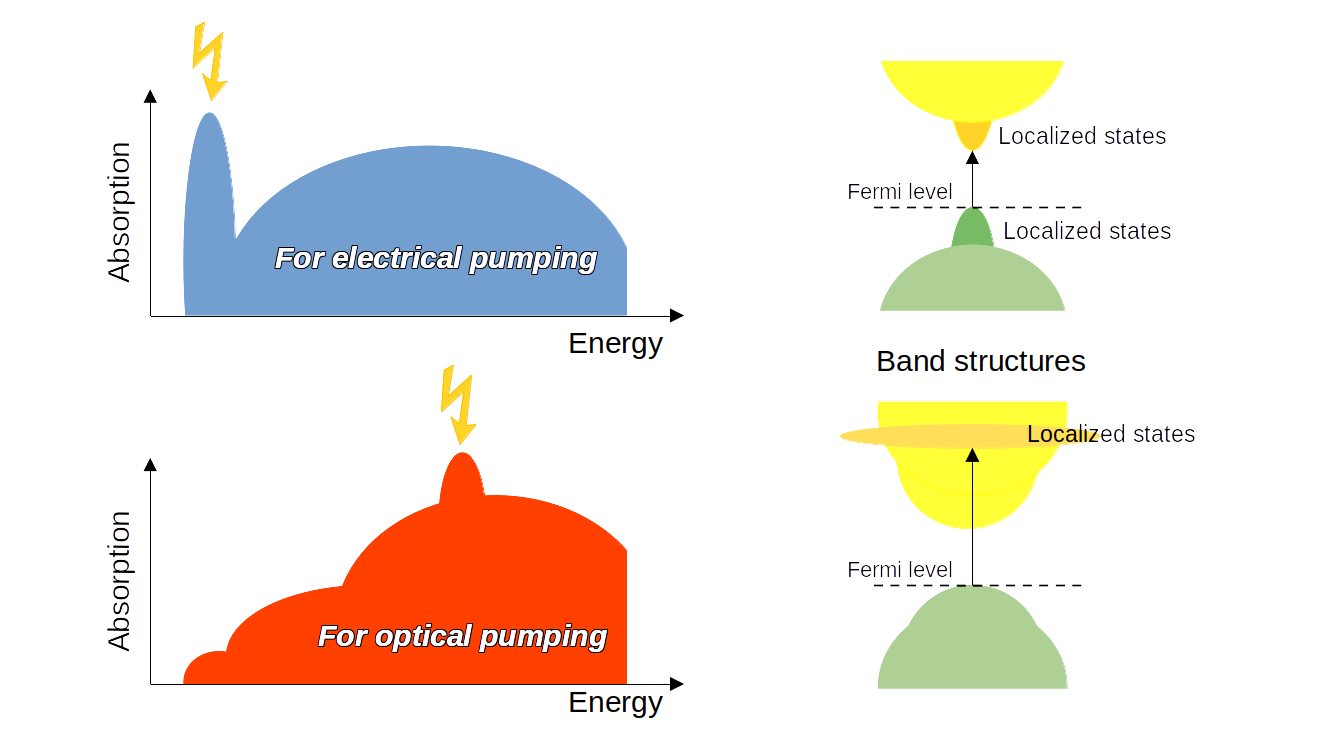}
    \caption{Schemes of the typical absorption spectra and corresponding band structures for electrically and optically pumped lasers.}
    \label{fig:F1}
\end{figure}

The conditions required for optimal performance of solar cells and lasers are opposed to each other. The former require semiconductors with weakly bound excitons, while the latter require a material with strongly bound excitons and short radiative relaxation times. We are interested here in lasers, and thus in exciton pairs with high binding energy, which are usually formed when electrons and holes are localised in a small region of the crystal forming Frenkel excitons \cite{F}. Localised electron-hole pairs can be achieved by a suitable doping pattern in the host material. This requirement is common to both optically and electrically pumped lasers. The difference between their respective modes of operation lies in the energetic position of the (localised) excited states, which have the shortest relaxation time within the conduction band manifold. In the case of electrically pumped lasers, the fast emission must come from deep conduction states that are accessible by the electrical charge injection. For optically pumped lasers, on the other hand, any strong emission from any conduction state can be appropriate. We highlight this rule in Figure~\ref{fig:F1}, which shows the (schematic) shapes of the absorption spectrum and related band structures for the two types of laser. It is easy to see that electrical pumping can only occur in an absorption spectrum with a high peak at the low-energy edge, i.e., with the corresponding band structure having localised states close to the band gap. Besides, electrical pumping requires some other physical conditions to be fulfilled, namely the existence of a direct gap (so that no phonons are involved in charge relaxation) and, more importantly, the existence of localised (i.e., with small radii) excitons. Note that the fulfilment of these conditions still does not guarantee that an electrically pumped laser constructed from the solid studied here will be directly suitable for practical applications, as we are not concerned with the electronic part of the devices. In particular, any discussion about efficiency is beyond the scope of this paper.

Returning to the perovskite system, the strongest emission of light from halide perovskites is expected from the Cs states located more than 4~eV above the conduction band minimum (CBM); according to the previous paragraph, these states make them good candidates for optical pumping \cite{PCCP}. The valence band maximum (VBM) in these systems consists mainly of states localised on the halide atoms, while the CBM consists mainly of states localised on the B cation (usually states of Pb or Sn). The states of the A-cations are known not to contribute to the lowest conduction bands, although some organic A-cations have been theoretically predicted to add their states not very far from the gap and give strong emission to other molecular cations \cite{Alicja}. 

In this paper we explore the idea that a halide perovskite with a peak at the onset of the absorption spectrum (and thus potentially useful for electrically pumped lasers) can be obtained  from currently available materials by a suitable doping scheme. In particular, a localised exciton near the gap can be created by adding a component with a small gap to a second one with a larger gap. As an example, one can dope ABBr$_3$ with ABI$_3$ or mix ABCl$_3$ with ABBr$_3$ \cite{CC}. Without losing generality, we have chosen here the CsPbX$_3$ perovskite, with X = Br as the host halide and X = I as the dopant. For simplicity, we restrict ourselves to the cubic phase. Our results show that such a configuration does indeed exhibit the desired feature at the onset of the absorption spectrum.

\section{Calculation details}

\begin{figure}[!ht]
    \centering
    \includegraphics[width=\textwidth]{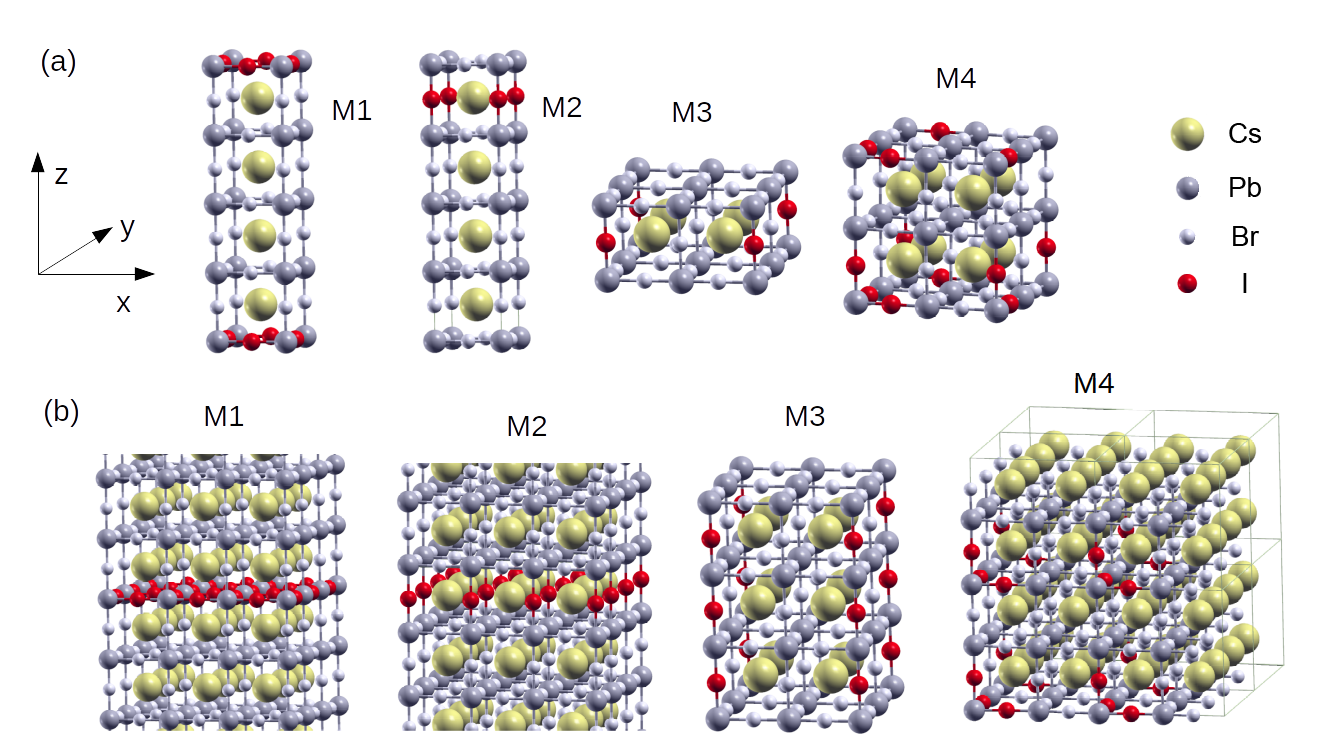}
    \caption{Atomic structures of the defective perovskite supercells. \textit a) M1 = (1,1,4)$a$, M2 = (1,1,4)$a$, M3 = (2,2,1)$a$ and M4 = (2,2,2)$a$ configurations. \textit b) Periodic repetitions of supercells M1, M2, M3 and M4.}
    \label{fig:F2}
\end{figure}

Several simulation supercells were constructed by periodically repeating the ABX$_3$ unit cell. The cases labelled M1 and M2 were constructed by stacking four elementary unit cells along the z-axis, resulting in supercells with lattice parameters (1,1,4)$a$, where $a$ is the lattice parameter of the perovskite unit cell. The M3 case is an in-plane repetition of the unit cell, with lattice parameters (2,2,1)$a$, and the M4 case is the bulky (2,2,2)$a$ supercell. In these defect-free supercells, some Br atoms were replaced by I atoms to simulate the defective perovskites. In M1 the dopant atoms were placed in the basal plane, while in M2 the dopants were placed in the BrCs plane of the undefective perovskite. In M2 the dopant I atoms form vertical I-Pb chains. Finally, the PbI$_3$ seeds are located within the M4 supercell. Figure~\ref{fig:F2} shows all the supercells considered in this paper.

The constructed supercells were first relaxed within the density functional theory (DFT) formalism \cite{Kohn, Sham}; the lattice constant was initially chosen as $a =$ 5.94~\AA, comparable to that computed by DFT methods for the defect-free CsPbBr$_3$ \cite{Lat-1, Lat-2, Lat-3}. In the relaxation process, we assumed that the epitaxial growth of the I-doped regions does not allow the planar change of the interatomic distances when the CsPbBr$_3$ crystal is used as a substrate. The DFT calculations were performed within the generalised gradient approximation (GGA) with the Perdew-Burke-Ernzerhoff \cite{PBE} parameterisation for the exchange-correlation functional using the Quantum Espresso package \cite{QE}, a code that uses a plane-wave basis set and pseudopotentials for the atomic cores. Norm-conserving pseudopotentials were used, with the energy cutoff for the plane waves set to 60 Ry. The first Brillouin zone (BZ) was sampled using uniform Monkhorst-Pack grids \cite{MP} with $k-$point meshes of 8$\times$8$\times$4 for M1 and M2, 4$\times$4$\times$8 for M3 and 4$\times$4$\times$4 for M4.

To draw the band structures we used the Wannier90 code \cite{Wan90}, which finds the maximally localised Wannier functions (MLWFs) using the DFT solutions \cite{Marzari, Wan-RMP}. The Hamiltonian was then interpolated into the MLWF basis and the band energies were associated with the projection of the corresponding Bloch functions onto the iodine-centred MLWFs. The special k-points along the band line directions are $\Gamma = (0,0,0)$, $X = (1,0,0)$, $M = (1,1,0)$, $Z = (0,0,1)$ and $R = (1,1,1)$.

The calculation and analysis of the optical spectrum was performed by solving the Bethe-Salpeter  equation (BSE) \cite{BSE-1} from the DFT eigenvalues and eigenvectors using the Yambo code \cite{Yambo-1, Yambo-2}, an implementation of the \textit{ab initio} many-body perturbation theory (\textit{ai}-MBPT). This combined formalism has been successfully applied to a large number of systems, including not only perovskite-based systems \cite{Motta,Xi}, but also other exotic solids such as topological insulators \cite{topological}. The BSE explicitly includes electron-hole correlations, such as those leading to the formation of excitons, in the Hamiltonian 
\begin{equation}
	H_{nn'{\bf k},mm'{\bf k'}} = (\varepsilon_{n{\bf k}}-\varepsilon_{n'{\bf k}})\delta_{nm}\delta_{n'm'}\delta_{\bf kk'} + (f_{n{\bf k}}-f_{n'{\bf k}}) \left[ \bar{V}_{nn'{\bf k},mm'{\bf k'}} - W_{nn'{\bf k},mm'{\bf k'}} \right],
	\label{eq:BSE_Hamiltonian}
\end{equation}
where $\varepsilon_{n\bf{k}}$ and $f_{n\bf{k}}$ denote the Kohn-Sham eigenvalues from the DFT and the Fermi-Dirac distribution function, respectively, 
\begin{equation}
	\bar{V}_{nn'{\bf k},mm'{\bf k'}} = \frac{1}{\Omega N_q} \sum_{\bf G \neq 0} \rho_{nn'}({\bf k},{\bf q}=0,{\bf G})
	\rho^{\ast}_{mm'}({\bf k'},{\bf q}=0,{\bf G}) v({\bf G}) 
	\label{eq:exchange_term}
\end{equation}
describes the exchange interaction and
\begin{equation}
	W_{nn'{\bf k},mm'{\bf k'}} = \frac{1}{\Omega N_q} \sum_{\bf G G'} \rho_{nm}({\bf k},{\bf q}={\bf k}-{\bf k'},{\bf G})
	\rho^{\ast}_{mm'}({\bf k'},{\bf q}={\bf k}-{\bf k'},{\bf 'G}) \epsilon^{-1}_{\bf GG'}v({\bf q}+{\bf G'}) 
	\label{eq:scattering_term}
\end{equation}
denotes the scattering term. In Eqs. \eqref{eq:exchange_term} and \eqref{eq:scattering_term}, $\Omega$ is the volume of the unit cell, $N_q$ is the number of points used to sample the BZ, 
\begin{equation}
	\rho_{cv\bf{k}}({\bf{q, G}}) =  \mel{c \bf{k}}{e^{i(\bf{q + G})\cdot\bf{r}}}{v, \bf{k-q}}, 
	\label{eq:rho}
\end{equation}
with $c$ and $v$ labelling the conduction and valence bands involved in the transitions of interest, respectively, and $\ket{n, \bf k}$ being the corresponding DFT eigenstates and $v({\bf{k + G}}) = \frac {4\pi}{|\bf{k + G}|^2}$ are the Fourier components of the unscreened Coulomb potential. In this work we have used the dipole approximation for the single transitions, so that
	\[ \rho_{cv\bf{k}}({\bf{q,G}}) = \lim_{\bf{q} \rightarrow 0} \mel{c\bf{k}}{\bf{r}}{v, \bf{k-q}} \equiv \mel{e}{\hat D}{h} \]

On the other hand, the $\epsilon^{-1}_{\bf GG'}$ term in \eqref{eq:scattering_term} corresponds to the components of the complex dielectric function. In general, the latter may computed from an MBPT approach such as the GW approximation \cite{GWA}. Unfortunately, this approach is well beyond our computational capabilities for the systems considered here. Instead, we have adopted a simplified scheme consisting on a rigid 0.5 eV upward shift of the entire conduction band. This so-called "scissor operator" approach is expected to slightly alter the quantitative results reported here without changing the overall qualitative behaviour. 

The diagonalisation of the BSE Hamiltonian \eqref{eq:BSE_Hamiltonian} yields the excitonic energies $E_\lambda$ and states $\ket{\lambda}$.These can be used to calculate the macroscopic dielectric function as \cite{Yambo-1}

\begin{equation}
	\epsilon_M(\omega) \equiv 1 - \lim_{{\bf q}\rightarrow 0} \frac{8\pi}{|{\bf q}|^2\Omega N_q}
	\sum_{nm {\bf k}} \sum_{n' m'{\bf k'}} \rho^{\ast}_{nn'{\bf k}}({\bf q},{\bf G}) \rho_{mm'{\bf k'}}({\bf q},{\bf G'})
	\sum_{\lambda} \frac{A^{\lambda}_{nn'{\bf k}}(A^{\lambda}_{mm'{\bf k'}})^{\ast}}{\omega - E_{\lambda}},
	\label{eq:epsilon_m}
\end{equation}
with 
	\[ A_{cv\bf{k}}^\lambda = \braket{c v \bf k}{\lambda}, \]
where $\ket{cv\bf k}$ denotes the DFT electron-hole state, i.e., the conduction band $c$ and valence band $v$ DFT states at the same $k$-point $\bf{k}$. Finally, from the macroscopic dielectric function, the absorption spectrum is calculated, which is proportional to the imaginary part of \eqref{eq:epsilon_m}. Note that, for a set of excitons $\{\lambda\}$, the dielectric function \eqref{eq:epsilon_m} is written as
\begin{equation}
	\epsilon_M(\omega) \propto \sum_{\lambda} \left| \sum_{e,h} A^{\lambda}_{e,h}\langle e|\hat{D}|h\rangle \right|^{2} \delta({E_{\lambda} -\hbar \omega}) \; = \; \sum_{\lambda} S^{\lambda} \; \delta({E_{\lambda} -\hbar \omega}),
 	\label{eq:strengths}
 \end{equation}
where $S^\lambda$ is the strength of the $\lambda$ exciton. 

Two approaches were used for the \textit{ai}-MBPT step from the DFT calculations. One of these was non-relativistic (NR). The other one was fully relativistic and explicitly included spin-orbit coupling (SOC) corrections; the inclusion of this effect is unavoidable, since the band composition (and hence the character of the excitons in the absorption spectrum) is strongly affected by relativistic effects \cite{SOC}. The k-point meshes used for the \textit{ai}-MBPT calculations were 18$\times$18$\times$8 for the M1 and M2 supercells, 12$\times$12$\times$18 for the M3 supercell and 12$\times$12$\times$12 for the M4 supercell, in order to obtain convergent absorption spectra within the Haydock diagonalisation scheme \cite{Lanczos}. An example of a test carried out to calibrate the number of $k-$points at the BZ is shown in Figure~S1 of the Supporting Information (SI) (file). The number of orbitals for the polarisation function was more than twice the number of electrons for each system. The plane waves were used up to energies of 10~Ry for the exchange components and 4~Ry for the screening and response block size. The electron-hole interaction space was built on 10~orbitals near the Fermi level for the M1, M2 and M3 cases and 20~orbitals for the M4 case. The electric field vector orientations were taken parallel to $[$100$]$, in the epitaxial plane and perpendicular to it (i.e., out-of-plane) along the z-axis $[$001$]$.

\section{Results and discussion}

\begin{figure}[!ht]
\sbox\twosubbox{%
  \resizebox{\dimexpr.9\textwidth-1em}{!}{%
    \includegraphics[height=3cm]{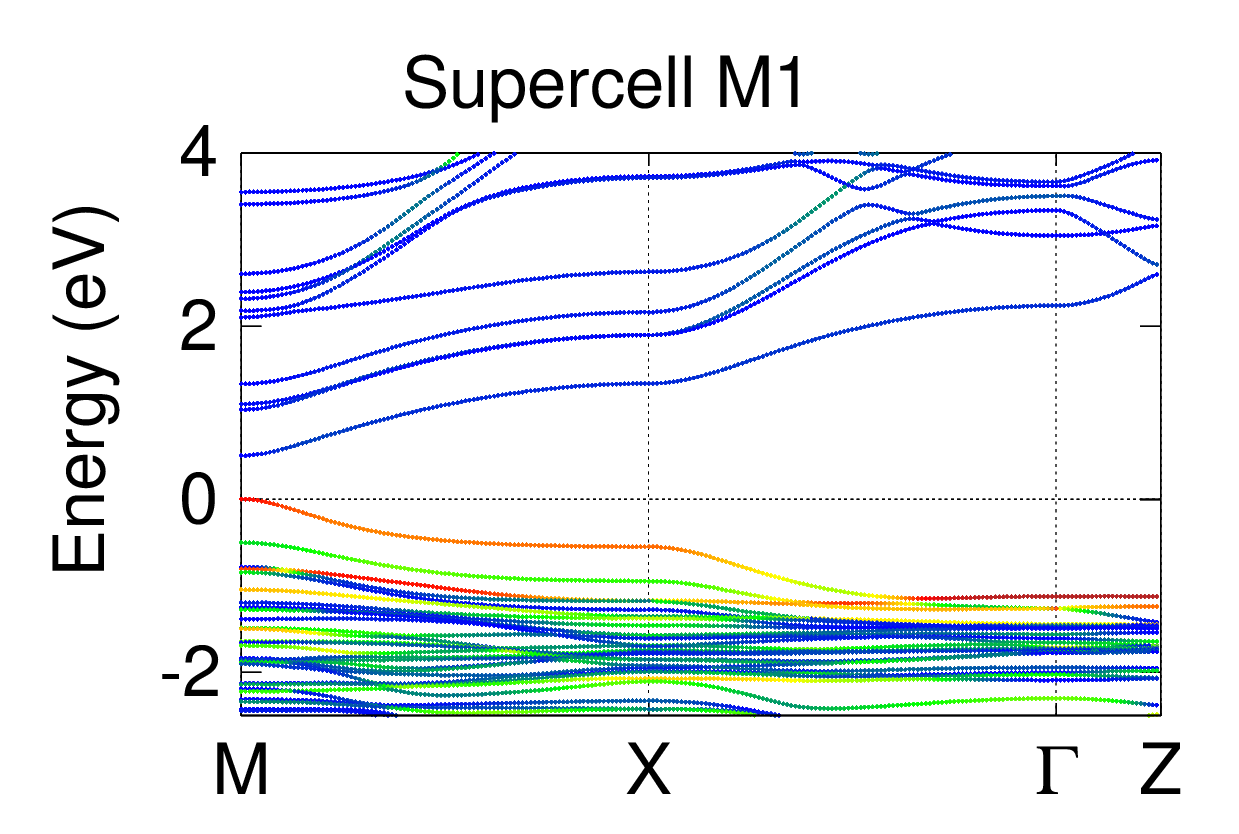}%
    \includegraphics[height=3cm]{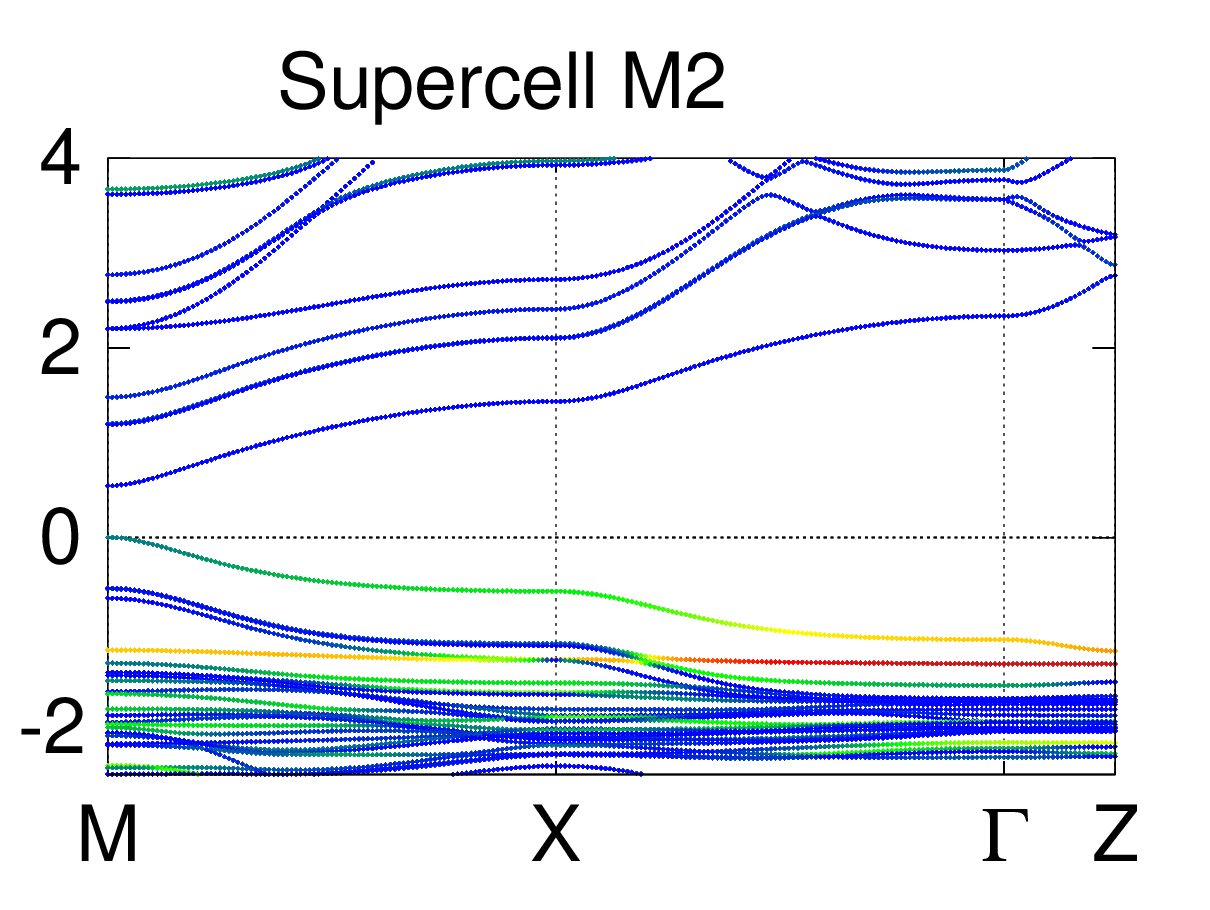}%
  }%
}
\setlength{\twosubht}{\ht\twosubbox}
    \includegraphics[height=\twosubht]{BW1.png}
    \quad
    \includegraphics[height=\twosubht]{BW2.png}%
\\
    \includegraphics[height=\twosubht]{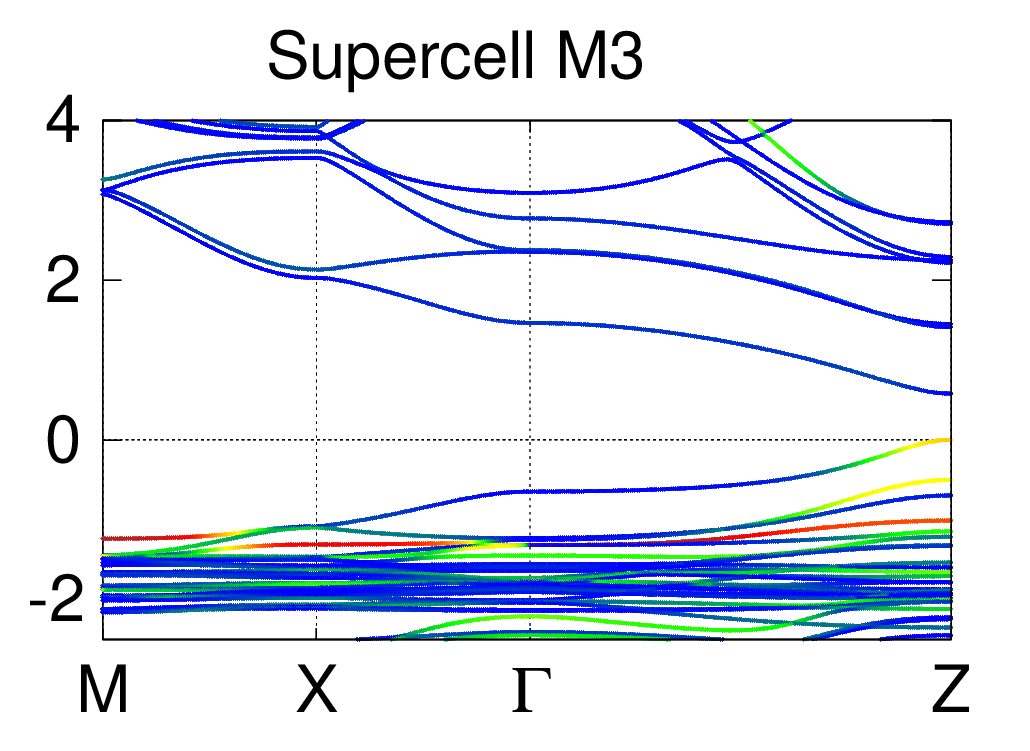}%
    \includegraphics[height=\twosubht]{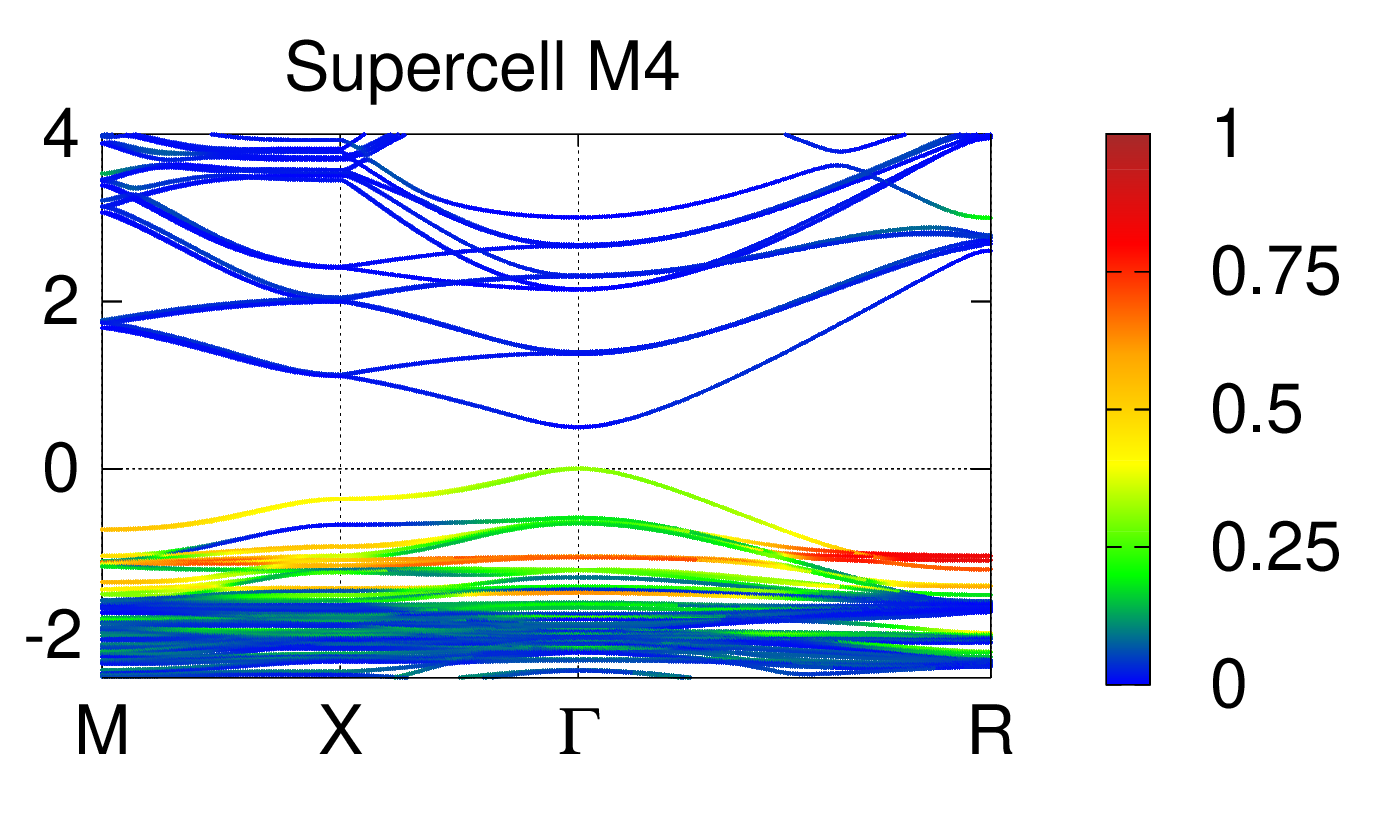}%

\caption{Band structures of the supercells considered in this work, calculated within the DFT including SOC. The colour code is common to all cases and represents the projection of the bands onto the maximally localised Wannier functions centred on the iodine atoms.}
\label{fig:F3}
\end{figure}

The band structures of the supercells obtained within the DFT with SOC are shown in Figure~\ref{fig:F3}. It is known that the geometry of the supercell defines the position of the fundamental band gap within the BZ. In our case, the M1 and M2 supercells are elongated along the z-direction and the gap appears at the $M-$symmetry point. On the other hand, the M3 supercell has a planar shape and the gap opens at $Z$, while the highly symmetric shape of the M4 supercell causes the gap to appear at $\Gamma$. In any case, the gaps are likely to be underestimated with respect to the experimental ones, if they were measured for these epitaxial structures. This is a well-known pathological and systematic effect inherent to DFT, which is due to the inability of any approximation for the correlation-exchange functional to account for the real self-interactions within the electron gas. This underestimation could be partially overcome by including pseudopotential self-interaction corrections (pSIC), which on the contrary widen the gaps, as discussed elsewhere \cite{Breath}. Unfortunately, to the best of our knowledge, there is no implementation of pSIC with SOC available. Therefore, the gap problem of DFT will be ignored in the following. After all, we are concerned here with the spectroscopic properties of the selected supercells, and not with the exact values of the band gaps.

The colour code in Fig.~\ref{fig:F3} shows the projection of the bands onto the MLWFs centred on the iodine atoms. As will be discussed below, these atoms strongly influence the optical performance of these systems. For the VBM the corresponding projections are 79\% for the M1 case and 10\% for M2 (both at the $M$ point), 38\% for M3 (at $Z$) and 31\% for M4 (at $\Gamma$). At first glance it might seem that the doping type of the M1 supercell would be preferable to the others, but this is mainly due to the higher content of I atoms per four formula units, i.e. two for M1 and one for M2 and M3; the M4 supercell contains 3 iodides per eight formula units. In fact, the M1 and M3 cases have equally well localised valence top states due to the symmetry of the doping pattern, which includes the Pb-I chains in 2D for M1 and 1D for M3. The PbI$_3$ grains in the M4 doping pattern interact less with each other because they are separated by the distance of an elementary formula unit. A larger separation would require many more atoms in the supercell; unfortunately, we cannot afford such a calculation using \textit{ai}-MBPT methods. Interestingly, the M2 case, which contains the Pb-I-Pb grains separated by three formula units along the z-axis and does not show a very significant contribution of I to the VBM, does exhibit a noticeable interaction between the iodides in the I-Cs plane. 

\begin{figure}[!ht]
    \centering
    \includegraphics[width=\textwidth]{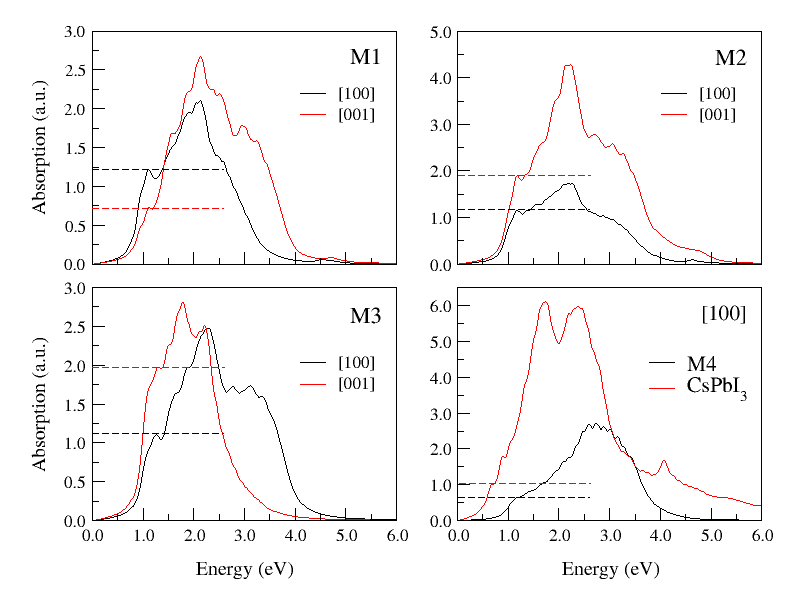}
    \caption{Absorption spectra obtained from the \textit{ab initio} Bethe-Salpeter equation for the supercells of Fig.~\ref{fig:F2}. Data for pure CsPbI$_3$ are included for comparison. The electric field is polarised along the $[100]$ (i.e. in-plane) and $[001]$ (out-of-plane) directions. Dotted lines indicate the height of the first prominent peak.}
    \label{fig:F4}
\end{figure}

To bring to light the optical properties of these systems, Figure~\ref{fig:F4} shows the absorption spectra calculated for all the studied supercells by solving the \textit{ab initio} BSE on top of the DFT+SOC solutions. As the low energy peaks are the most relevant for the following discussion, the dotted lines in each plot highlight the heights of these peaks. For the anisotropic supercells (M1, M2 and M3) the calculations were performed with two orientations of the electric field vector, namely in-plane [110] and out-of-plane [001] orientations with respect to the doping pattern. The M4 supercell is symmetric with respect to all three Cartesian axes; the same symmetry characterises the elementary cells of the pure compounds. For comparison, the absorption spectrum of pure CsPbI$_3$ (from ref. \cite{PCCP}) is added to the Bethe-Salpeter spectrum in this case. For completeness, the corresponding spectra obtained from NR calculations for the four supercells are shown in Figure~S2 in the SI. One can quickly see that the shapes of the curves calculated with and without SOC are very different due to the orbital splitting of the large and small spinors. Also for comparison purposes, we include in the SI the absorption spectra of Fig.~\ref{fig:F4} normalised to the height of the main peak in each case (Figure~S3). In what follows, we will further analyse the results obtained with the relativistic methods.

For both M4 and pure CPbI$_3$, the height of the first absorption peaks is around 20\% of the maximum of the absorption curves, while the anisotropic doping results in much higher first absorption peaks, reaching 60-70\% of the maximum peak. For this reason, we will omit the study of the M4 configuration, as it is unlikely to give optimal optical properties. On the other hand, the absorption spectra of the M1 and M3 supercells show that the orientation of the electric field should be within the Pb-I$_2$ plane (for M1) and along the Pb-I chain (for M3) in order to increase the low energy edge. This means that the crystal orientation should obey the rule that the Pb-I bonds of the doping pattern are perpendicular to the absorbed/emitted light beam. 

As for the M2 doping pattern, an electric field polarised along the Pb-I-Pb bonds (i.e. [001]) yields a strong absorption peak, while a much more modest maximum appears for the perpendicular [100] orientation (i.e., within the Cs-I plane). In other words, the intensity of the first peak is stronger when the light beam is perpendicular to the Pb-I-Pb bonds, confirming our previous observation for the M1 and M3 cases. It should be noted that the absolute and normalised absorbance curves for the M2 and M3 cases do not exchange the heights of the edge peaks for the two calculated polarisations of the electric field.

\begin{figure}
	\centering
	\includegraphics[width=0.49\textwidth]{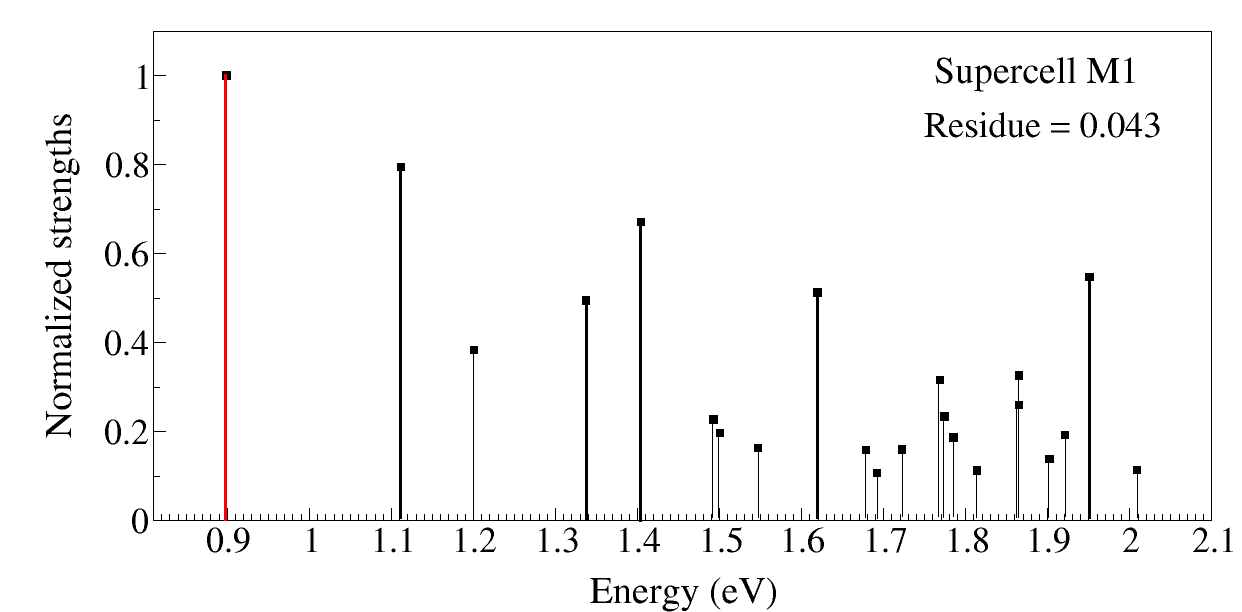} 
	\includegraphics[width=0.49\textwidth]{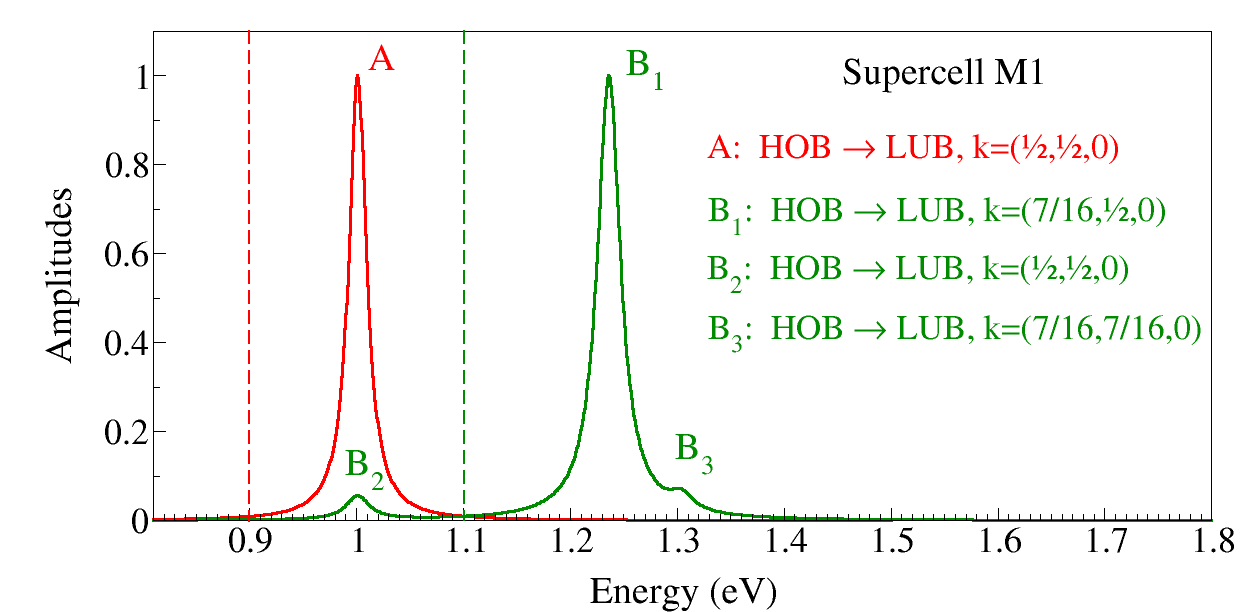} \\
	\includegraphics[width=0.49\textwidth]{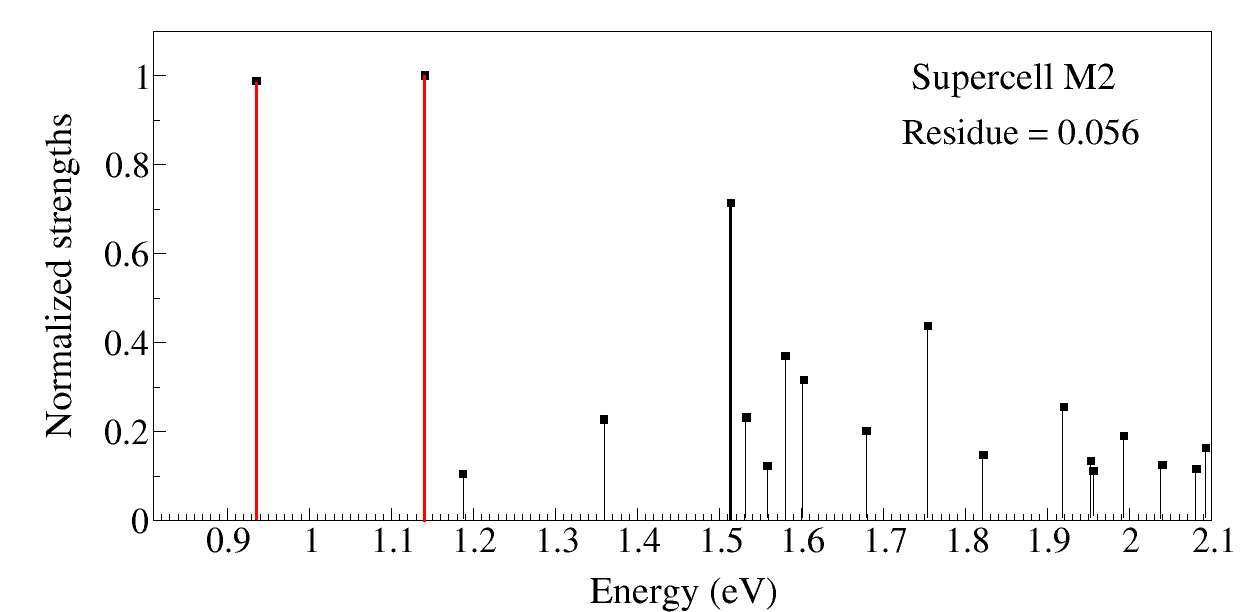} 
	\includegraphics[width=0.49\textwidth]{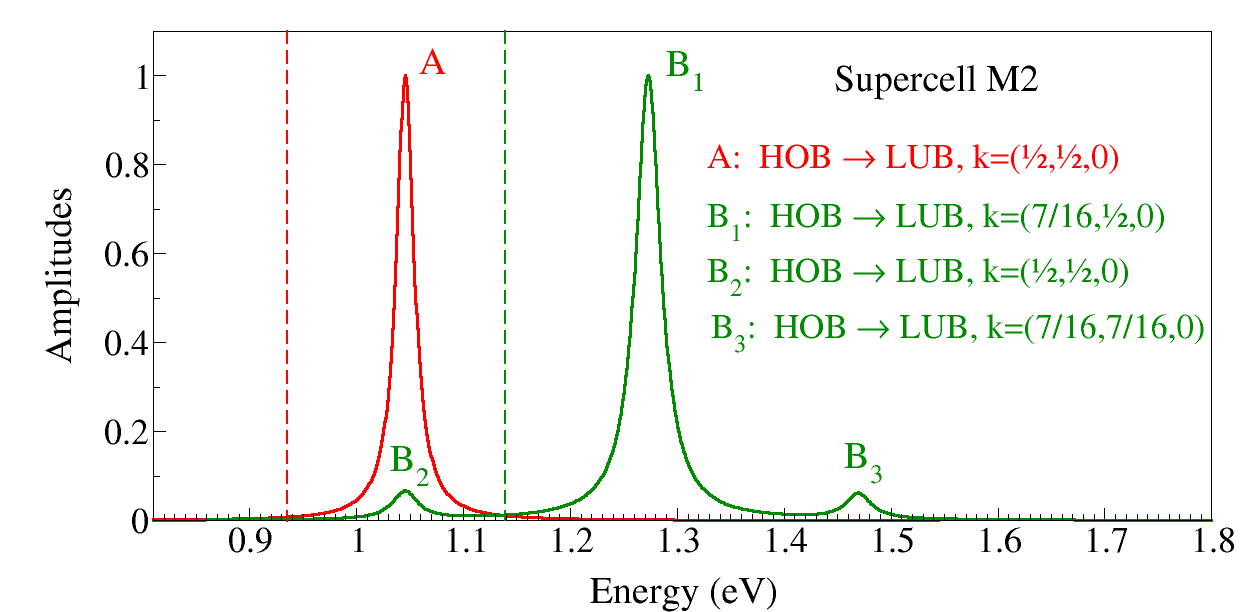} \\
	\includegraphics[width=0.49\textwidth]{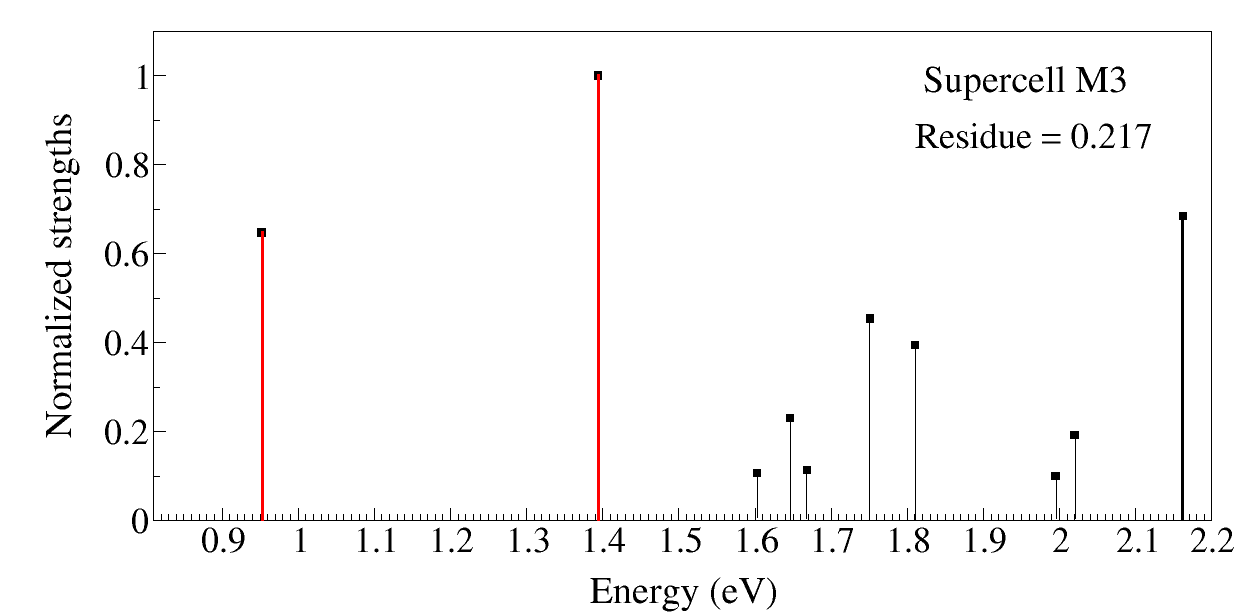} 
	\includegraphics[width=0.49\textwidth]{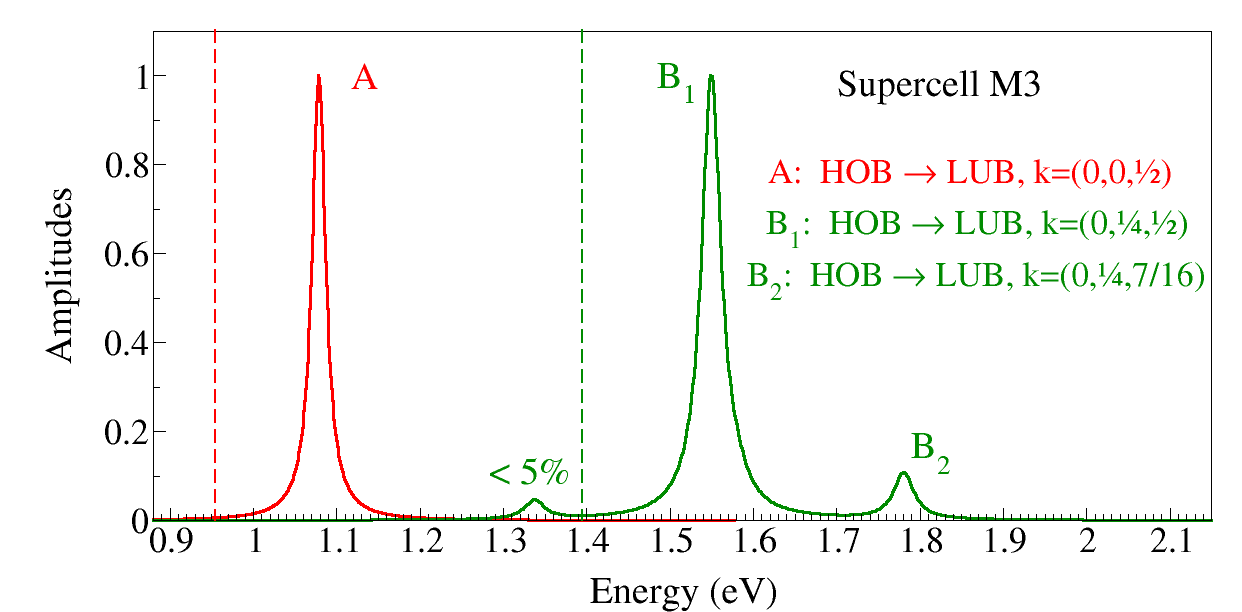} \\
	\includegraphics[width=0.49\textwidth]{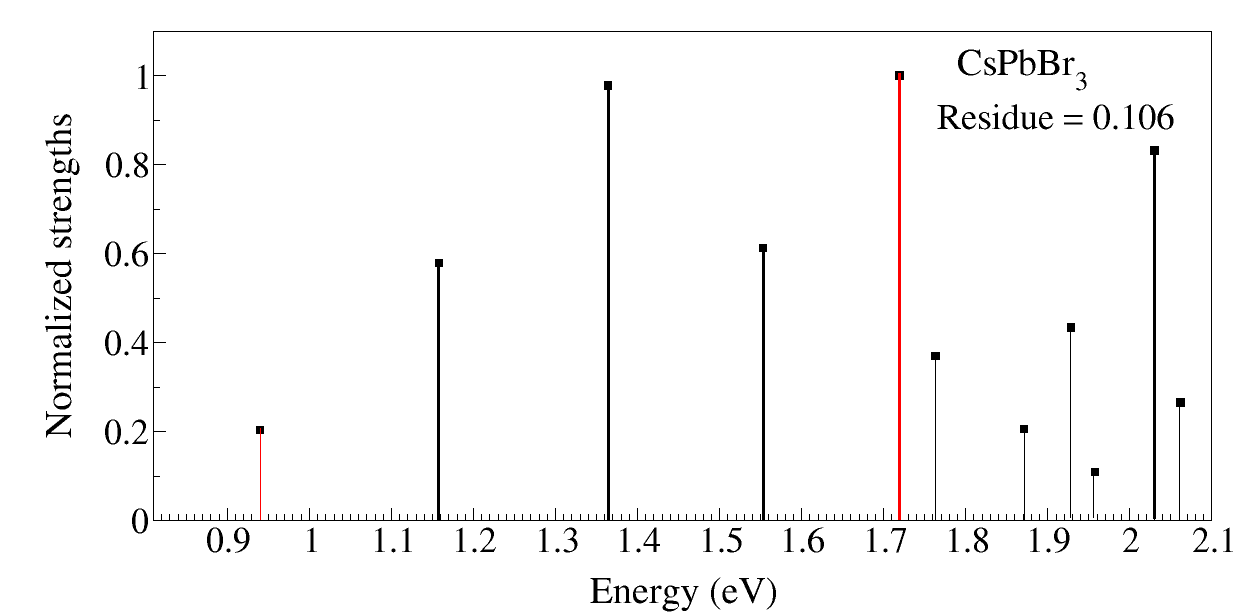} 
	\includegraphics[width=0.49\textwidth]{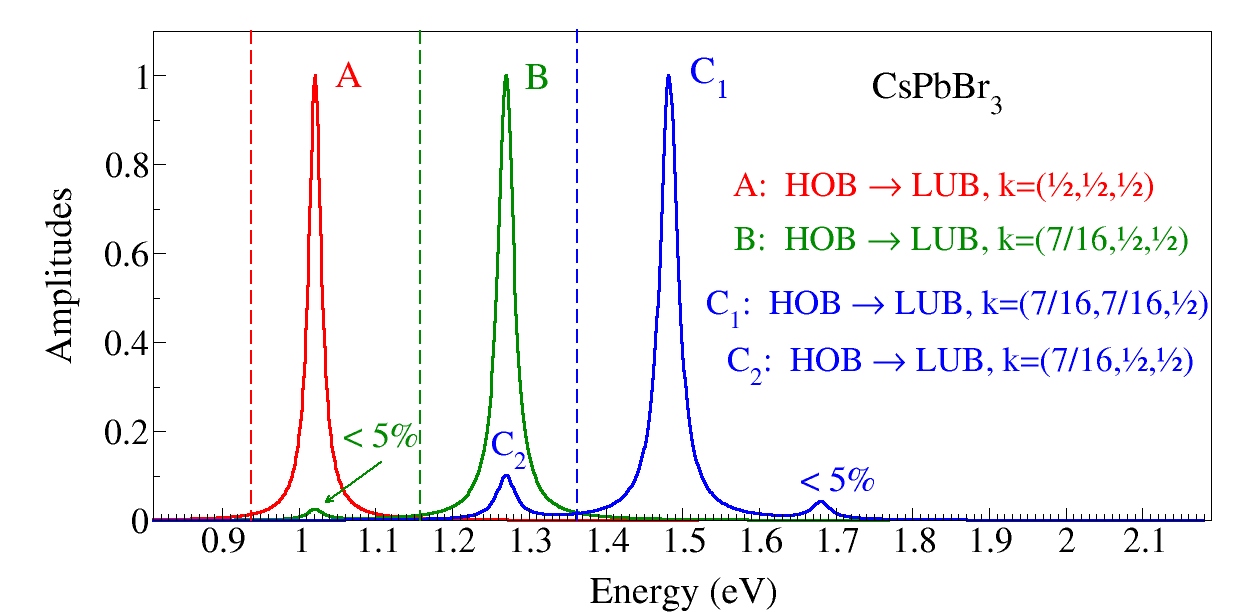} \\
	\includegraphics[width=0.49\textwidth]{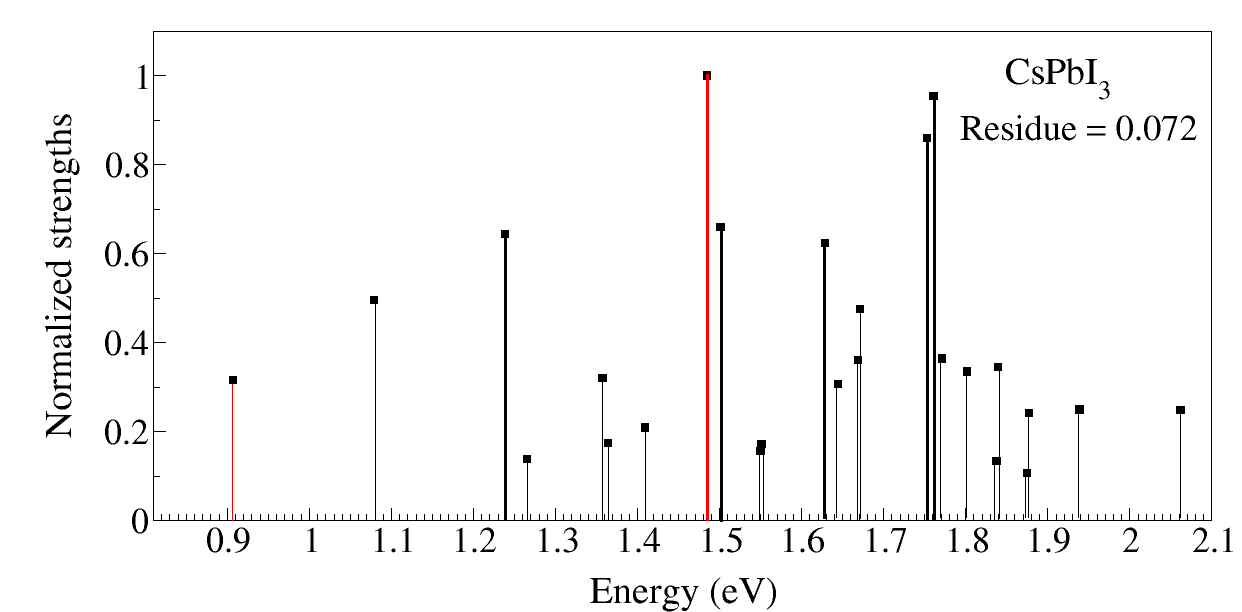} 
	\includegraphics[width=0.49\textwidth]{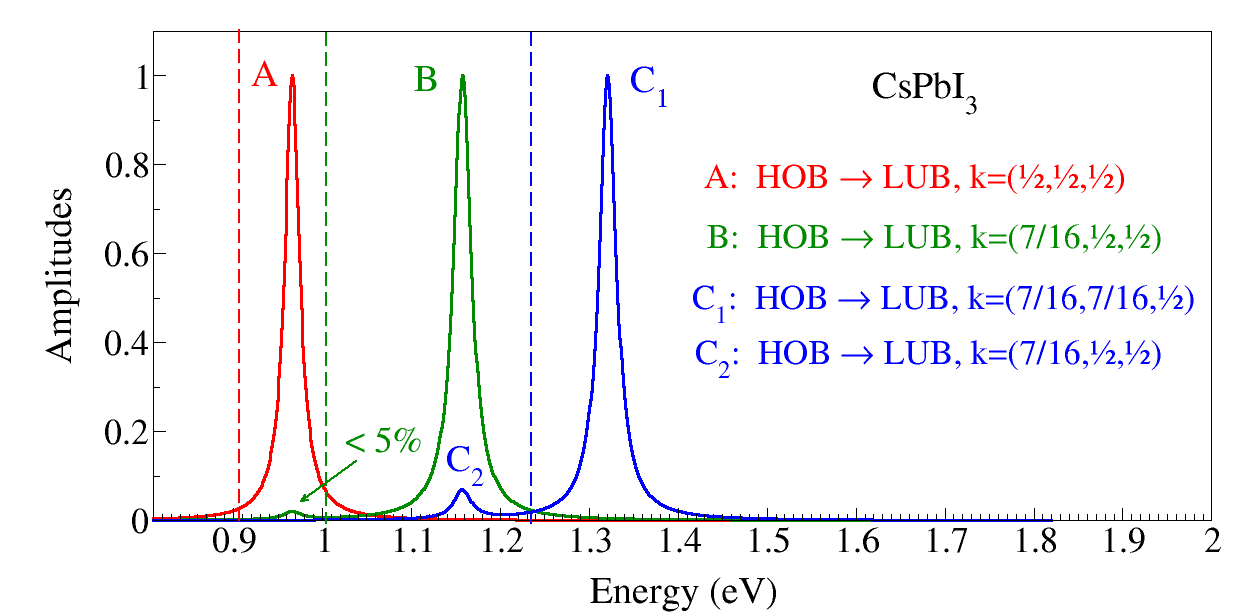}
	\caption{Left: Normalised strengths [$\bar S^\lambda$ in Eq.~\eqref{eq:strengths}] of the excitons contributing to the absorption spectra in Figs.~\ref{fig:F4}, as well as for the parent undoped perovskites CsPbBr$_3$ and CsPbI$_3$. The red lines highlight the lowest energy and the most intense excitons in each case. Only excitons with normalised strenghts above 5$\%$ are shown. Right: Amplitudes [$A_{e,h}^\lambda$ in Eq.~\eqref{eq:strengths}] of the lowest energy and the most intense excitons, centred on the DFT energies + 0.5~eV (see the main text for details). Capital letters denote the corresponding hole (HOB = highest occupied band) to electron (LUB = lowest unoccupied band) transitions and the $k$-point in the BZ at which they occur.}
	\label{fig:F5}
\end{figure}

According to \eqref{eq:strengths}, the absorption spectrum may be regarded as the density of states of the exciton strengths. Implications of the spectra in Fig.~\ref{fig:F4} in terms of band transitions require the analysis of the absorption spectra in terms of the strengths [$S^\lambda$ in Eq.~\eqref{eq:strengths}] of the excitons arising as solutions of the BSE for each configuration. In particular, Figures~\ref{fig:F5} left show the excitonic strengths $\bar S^\lambda$ normalised to the maximum one in the spectrum of each defective supercell and pure compound studied in this paper. The first, lowest energy exciton and the strongest one are highlighted in red color, while those excitons with normalised strengths larger than 0.5 appear in black. Excitons with strengths less than 0.1 are neglected. Note that in order to obtain the absolute strengths $S^{\lambda}$, one should to multiply the plotted values by the value of the residue $\dfrac {8\pi}{|{\bf q}|^2\Omega N_q}$, according to \eqref{eq:epsilon_m}. These residues are shown in the corresponding strength plots. It is interesting to note that, for both M1 and M2 configurations, the most intense exciton is also the one with the lowest energy. For M3, on the other hand, a relatively intense exciton (but not the most intense) is still observed at the absorption edge. These facts contrast with the situation in the parent pure perovskites CsPbBr$_3$ and CsPbI$_3$, whose most intense exciton does not appear at the onset of the respective spectra, and whose initial exciton is comparatively small. 

To the right of each exciton strength plot we show the amplitudes $A^{\lambda}$ for the first low-energy excitons and the corresponding decomposition into dipole transitions, i.e., the valence and conduction transitions and the $k$-point in the BZ zone where they occur. As a consequence of the aforementioned rigid opening of the gap, the amplitudes are centred at the DFT energies shifted 0.5 eV upwards, while the actual exciton energies are red-shifted due to the electron-hole interactions, which are taken into account at the BSE level but not within DFT. For this reason, we highlight the actual exciton energies in the amplitude plots as vertical dashed lines with similar codes. The capital letters label the valence and conduction states and the $k$-point at which the corresponding transition takes place. It is interesting to notice from these amplitude plots that we did not find any transitions (contributing at least 5$\%$, as already mentioned) involving states below two degenerate spinors at the highes occupied band (HOB) and above those at the lowest unoccupied band (LUB). 

\begin{figure}
	\centering
	\includegraphics[width=0.75\textwidth]{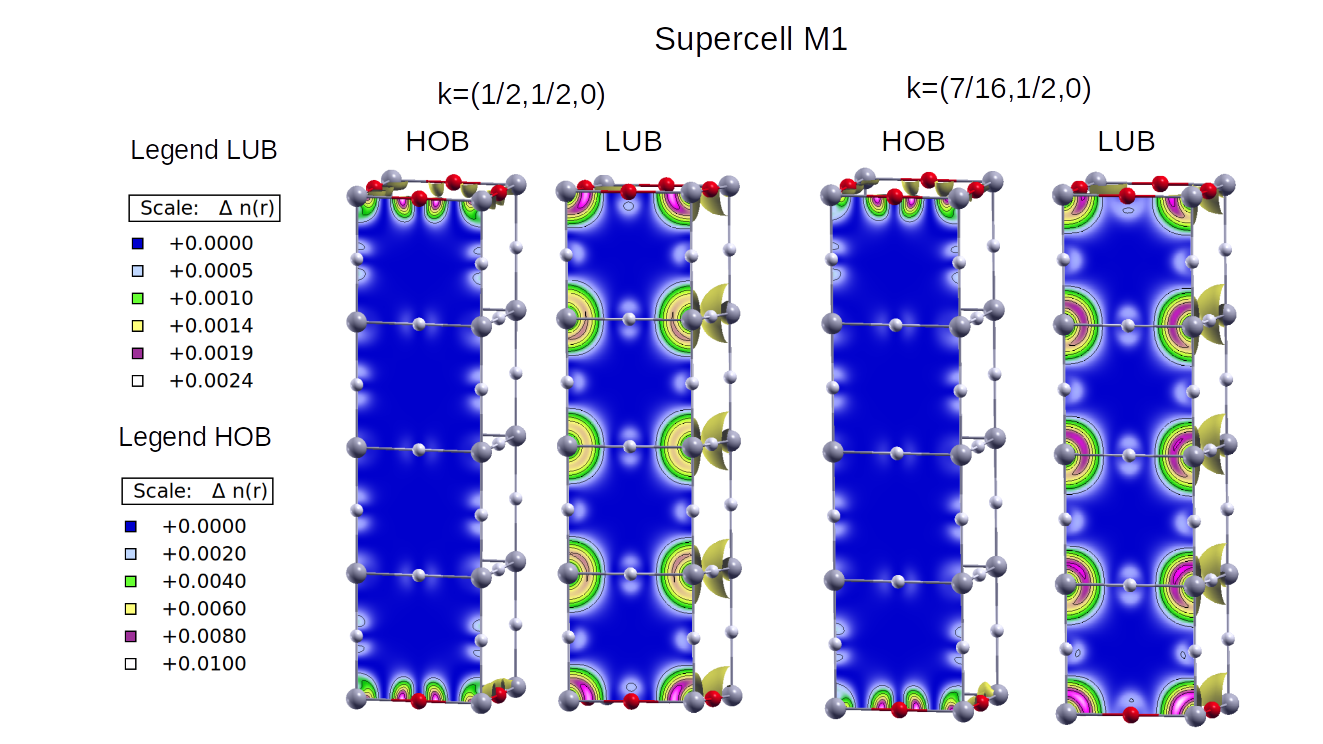}
	\caption{Bloch maps for electrons located at the highest occupied band (HOB) and lowest unoccupied band (LUB) for the lowest energy exciton [at ${\bf{k}} = \left(\frac 12, \frac 12, 0\right)$, left] and the most intense one [at ${\bf{k}} = \left( \frac 7{16}, \frac 12, 0\right)$, right] of the M1 configuration.}
	\label{fig:F6}
\end{figure}

\begin{figure}
	\centering
	\includegraphics[width=0.75\textwidth]{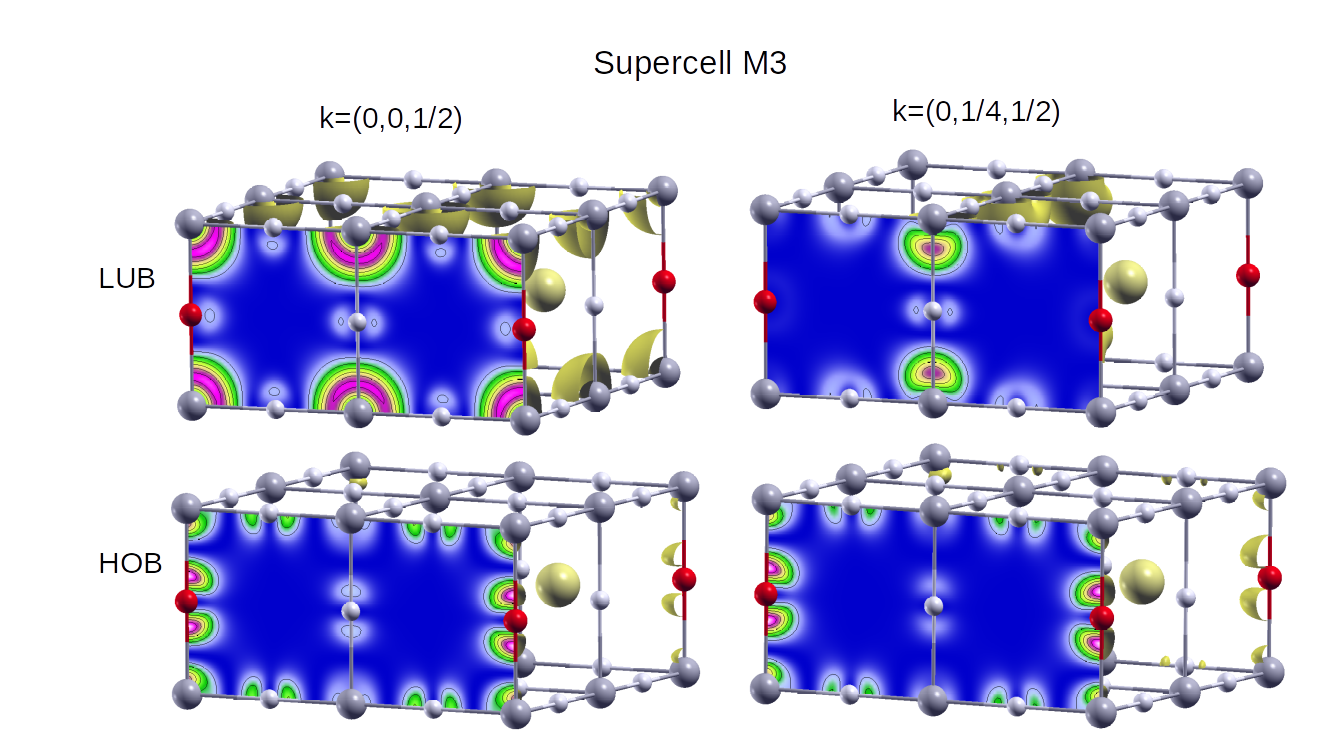}
	\caption{Idem for M3. In this case, the lowest energy (left) and most intense excitons (right) appear at ${\bf{k}} = \left(0, 0, \frac 12\right)$ and ${\bf{k}} = \left(0, \frac 14, \frac 12\right)$, respectively. The color code is the same as for Fig.~\ref{fig:F6}}
	\label{fig:F7}
\end{figure}

We can gain additional insight into the excitonic behaviour of the defective perovskites by plotting the Bloch maps [i.e. $\abs{\psi_{n,\bf{k}}(\bf{r})}^2$, where $n, \bf{k}$ denote band and $k$-point, respectively] for the states involved in the transitions reported in Figs.~\ref{fig:F5} right for the defective configurations. These plots help to understand the preferential location of electrons in each state and thus to estimate the position of the excitons in the direct space. Figures~\ref{fig:F6} and \ref{fig:F7} show the Bloch maps for the lowest energy (left) and the most intense (right) excitons for configurations M1 and M3, respectively. They correspond to iodine doping patterns which are a kind of 2D and 1D structures embedded in the 3D host.
We omit the M2 supercell from further analysis, because its shape is the same as for M1 supercells and the amplitudes in Fig.~\ref{fig:F5} show for M2 the same composition in terms of band-to-band optical transitions as for M1.

Fig.~\ref{fig:F6} evidence that electrons at the HOB locate preferentially near iodine atoms, with some appreciable density near Pb atoms too. At the LUB, on the other hand, electrons are more likely to locate at Pb sites, so transitions between iodine and lead atoms are expected for this configuration. The same effect holds for both the most intense and the lowest energy excitons. As for M3 (Fig.~\ref{fig:F7}), the HOB is highly hybridised, and the contributions are different for the lowest energy (left) and the most intense (right) excitons. In the former case, electrons at HOB are located around iodine, with some contributions from lead and bromide atoms, while for LUB they surround lead. Interestingly, the most intense exciton corresponds to transitions between iodine (there is no charge density around bromide at HOB) and lead atoms, as shown in Fig.~\ref{fig:F7} right. 

\begin{figure}[!ht]
    \centering
    \includegraphics[width=\textwidth]{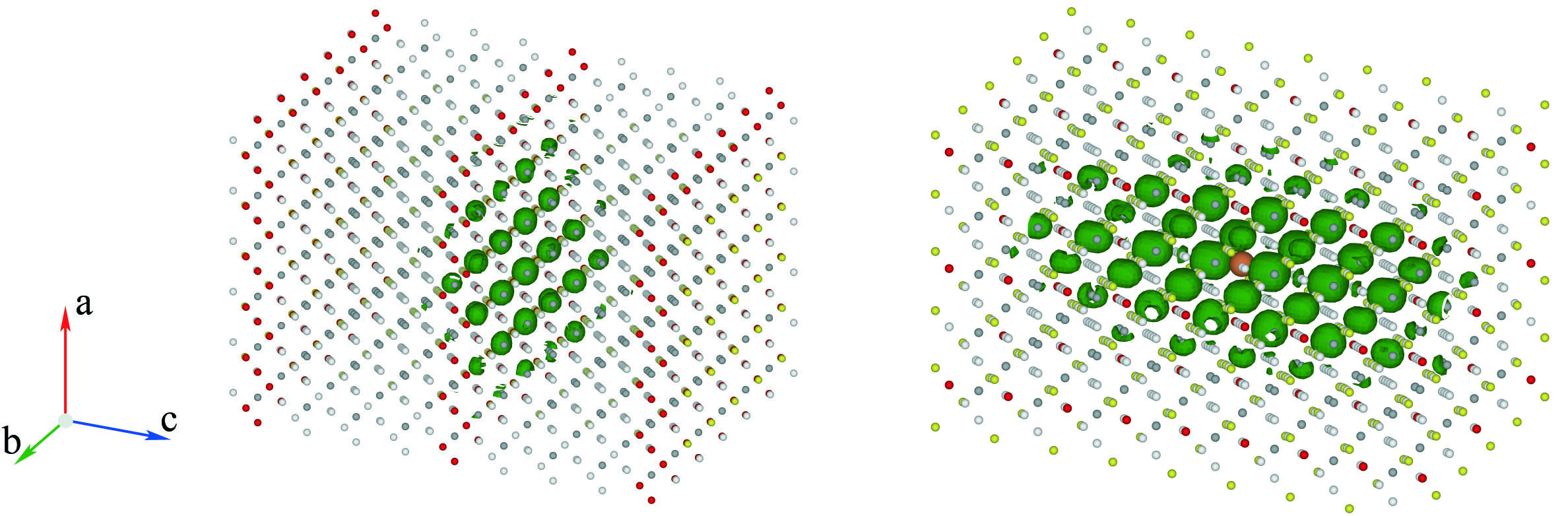} \\
    \caption{Isoprobability surfaces ($p = 2.5 \cdot 10^{-5}$) for finding an electron at a fixed position of the hole for the most intense exciton of configurations M1 (left) and M3 (right). In both cases, the hole has been located next to an iodine atom, as these make a prominent contribution to the VBT.}
    \label{fig:F8}
\end{figure}

We can now visualise the excitons mentioned above. To this end, Figure~\ref{fig:F8} displays the surfaces with equal probability of finding an electron for a given position of the hole (shown as a golden sphere in each case) for the most intense exciton of configurations M1 and M3. In both cases, the hole has been located near iodide atoms, which give prominent contributions to the VBT --in particular the maximum for M1. We notice that these excitons extend over five to seven unit cells, which is in good agreement with the (scarce) experimental values for excitons radii in perovskite systems ($r \sim$ 2.3 nm \cite{size1}). These exciton sizes are important for a potential use of these defective systems as components of electrically pumped lasers, as we will discuss below. 

Indeed, the electrical pumping of a laser is achieved by injecting charge into the bands in a very narrow energy range close to the VBM and the CBM. On the other hand, efficient radiative relaxation is possible in systems that obey many rules, which have been mentioned in the introduction. One of them is that the band gap must be direct, because in such a case the radiative transitions do not involve phonons that increase the excitonic lifetimes. All the supercells considered in this paper were found to have direct gaps. A second condition for efficient emission in lasers is a small size of the exciton radii. The radius of an exciton is inversely proportional to the expansion in direct space of the Bloch functions for the bands involved in the optical transitions. One would therefore expect that the optimal situation would be for the VBM to be composed mainly of iodide atomic orbitals. This is indeed the case for M1, as mentioned above, and to a lesser extent for M3. Note also that these excitons have relatively small sizes, as shown in Fig.~\ref{fig:F8}. Finally, the third requirement for a system to be a good optically active material is that it must have a prominent peak at the low energy edge of the absorption spectrum. The data in Figs.~\ref{fig:F4} and \ref{fig:F5} do indeed show that the defective configurations M1 to M3 exhibit excitonic peaks at the onset of their respective absorption spectra, and that these correspond to relatively intense excitons with properties that are in principle suitable for electrical pumping in lasers. We remark that these arguments do not allow us to conclude anything about the possible efficiency of a hypothetical laser containing these perovskites, as we mentioned in the Introduction section. 

To conclude, for the sake of completeness, the absolute absorption spectra obtained from the NR calculations are presented in Figure~S4 of SI. It should be noted, however, that these spectra should not be used as a reference for the experimental data, since the spinor splitting that leads to the fingerprint of the absorption curves is missing.

\section{Conclusions}

Despite the fact that many optoelectronic devices have been built using halide perovskites, including optically pumped lasers of all types, the construction of the electrically pumped perovskite laser still remains a challenge. In this theoretical work, we propose a special type of halide mixing that results in an absorption spectrum highly favoured for low-energy edge emission. The proposed architectures use the perovskite with the shortest lattice constant as the host material and the material with the longest lattice constant as a one-atom thick interlayer that localises the excitons spatially in the crystal and energetically in the spectrum. The study was carried out using \textit{ai}-MBPT methods implemented in the Yambo package to calculate the absorption curves by solving the Bethe-Salpeter equation. Experimental realisation of the structures presented would be possible using recently well-developed epitaxial techniques.

\section{Acknowledgements}
MW acknowledges financial support from Grant No. 2019/33/B/ST8/02105 funded by the National Science Centre of Poland. JJM acknowledges financial support from Grant No. PID2020-112936GB-I00 funded by MCIN/AEI/10.13039/501100011033) and from Grant No. IB20079 funded by Junta de Extremadura (Spain) and by ``ERDF A way of  making Europe''. The calculations have been supported by the National Science Centre of Poland, grant no. 2019/33/B/ST8/02105, and were performed using the PL-GRID infrastructure.

\bibliography{bibliography}
\bibliographystyle{rsc}

\balance
\end{document}